\documentclass[pdflatex,sn-vancouver-num]{sn-jnl}

\usepackage{graphicx}%
\usepackage{multirow}%
\usepackage{amsmath,amssymb,amsfonts}%
\usepackage{amsthm}%
\usepackage{mathrsfs}%
\usepackage[title]{appendix}%
\usepackage{xcolor}%
\usepackage{textcomp}%
\usepackage{manyfoot}%
\usepackage{booktabs}%
\usepackage{algorithm}%
\usepackage{algorithmicx}%
\usepackage{algpseudocode}%
\usepackage{setspace}   
\theoremstyle{thmstyleone}%
\theoremstyle{thmstyletwo}%

\theoremstyle{thmstylethree}%

\begin{document}

\title[Article Title]{Fully coherent short wavelength free-electron laser driven by a single sub-microjoule seed}


\author[1,2,3]{\fnm{Lanpeng} \sur{Ni}}
\author*[1,2]{\fnm{Zheng} \sur{Qi}}\email{qiz@sari.ac.cn}
\author[1]{\fnm{Xingtao} \sur{Wang}}
\author[1]{\fnm{Weiyi} \sur{Yin}}
\author[1,2]{\fnm{Zhen} \sur{Wang}}
\author[1,2]{\fnm{Kaiqing} \sur{Zhang}}
\author[1]{\fnm{Zhangfeng} \sur{Gao}}
\author[1]{\fnm{Nanshun} \sur{Huang}}
\author[1]{\fnm{Hanxiang} \sur{Yang}}
\author[1]{\fnm{Hang} \sur{Luo}}
\author[1,2]{\fnm{Si} \sur{Chen}}
\author[4]{\fnm{Junhao} \sur{Liu}}
\author[3]{\fnm{Yaozong} \sur{Xiao}}
\author[5]{\fnm{Lingjun} \sur{Tu}}
\author[5]{\fnm{Xiaofan} \sur{Wang}}
\author[1]{\fnm{Cheng} \sur{Yu}}
\author[1]{\fnm{Yongmei} \sur{Wen}}
\author[1]{\fnm{Fei} \sur{Gao}}
\author[1]{\fnm{Yangyang} \sur{Lei}}
\author[1]{\fnm{Jian} \sur{Chen}}
\author[1]{\fnm{Huan} \sur{Zhao}}
\author[1,2]{\fnm{Xiaoqing} \sur{Liu}}
\author[1]{\fnm{Lie} \sur{Feng}}
\author[1]{\fnm{Yanyan} \sur{Zhu}}
\author[1]{\fnm{Jiaqiang} \sur{Xu}}
\author[1]{\fnm{Liping} \sur{Sun}}
\author[1,2]{\fnm{Wenyan} \sur{Zhang}}
\author[1]{\fnm{Yue} \sur{Wang}}
\author[1]{\fnm{Taihe} \sur{Lan}}
\author[1]{\fnm{Ximing} \sur{Zhang}}
\author[1,2]{\fnm{Bin} \sur{Li}}
\author*[1,2]{\fnm{Tao} \sur{Liu}}\email{liut@sari.ac.cn}
\author*[1,2]{\fnm{Chao} \sur{Feng}}\email{fengc@sari.ac.cn}
\author[6]{\fnm{Dao} \sur{Xiang}}
\author[1,2]{\fnm{Bo} \sur{Liu}}
\author*[1,2]{\fnm{Zhentang} \sur{Zhao}}\email{zhaozt@sari.ac.cn}

\affil[1]{\orgname{Shanghai Advanced Research Institute, Chinese Academy of Sciences}, \city{Shanghai}, \postcode{201210}, \country{China}}
\affil[2]{\orgname{University of the Chinese Academy of Sciences}, \city{Beijing}, \postcode{100049}, \country{China}}
\affil[3]{\orgname{Zhangjiang Laboratory}, \city{Shanghai}, \postcode{201204}, \country{China}}
\affil[4]{\orgname{ShanghaiTech University}, \city{Shanghai}, \postcode{201210}, \country{China}}
\affil[5]{\orgname{Institute of Advanced Light Source Facilities}, \city{Shenzhen}, \postcode{518107}, \country{China}}
\affil[6]{\orgname{School of Physics and Astronomy, Shanghai Jiao Tong University}, \city{Shanghai}, \postcode{200240}, \country{China}}


\abstract{
\setstretch{1.2}

High-repetition-rate, fully coherent extreme-ultraviolet (EUV) and X-ray free-electron lasers (FELs) are essential for advanced time-resolved ultrafast spectroscopies. While external seeding serves as the standard technique to achieve precise temporal coherence, conventional methods demand hundred-megawatt peak-power laser systems. Furthermore, advanced configurations like echo-enabled harmonic generation (EEHG) introduce the severe complexities of dual-laser synchronization. Together, these requirements fundamentally restrict operations to kilohertz repetition rates and compromise overall system stability. Here, we experimentally demonstrate a fully coherent EEHG-FEL driven by a single, sub-microjoule seed laser. By employing a direct-amplification enabled harmonic generation technique, we utilize an initial 0.4 µJ (2 MW peak power) ultraviolet seed to directly drive coherent lasing at nanometer wavelengths. By eliminating the need for extreme peak powers and multiple synchronized lasers, this approach significantly simplifies the seeding architecture and provides a practical and robust pathway toward megahertz-class, fully coherent EUV and X-ray light sources.
}

\maketitle

\section*{Introduction}\label{sec1}

X-ray free-electron lasers (XFELs) have enabled femtosecond and sub-femtosecond time-resolved studies with atomic spatial resolution, driving significant advances in quantum materials and biomolecular dynamics \cite{pellegrini2016physics,mcneil2010x,barletta2010free,bostedt2016linac,huang2021features,feng2018review}. Presently, most XFEL facilities operate in the self-amplified spontaneous emission (SASE) regime \cite{kondratenko1980,bonifacio1984,ackermann2007operation,emma2010first,ishikawa2012compact,kang2017hard,prat2020compact,decking2020mhz}. Because SASE radiation initiates from electron beam shot noise, it inherently suffers from limited longitudinal coherence and severe pulse-to-pulse intensity fluctuations \cite{kondratenko1980,bonifacio1984}. External seeding methods, such as high-gain harmonic generation (HGHG) and echo-enabled harmonic generation (EEHG), overcome these limitations by imprinting the full coherence properties of an external laser onto the electron beam, which are subsequently transferred to the FEL photon beam \cite{yu1991generation,yu2000high,allaria2012highly,stupakov2009,xiang2009echo,zhao2012first,rebernik2019coherent}. In addition to inheriting the exceptional coherence of the seed, these FEL pulses are naturally synchronized to the external laser system, a critical requirement for precision pump-probe experiments. Notably, EEHG has demonstrated fully coherent FEL lasing down to 5.9 nm and coherent emission at harmonics ranging from the 84th to the 101st \cite{rebernik2019coherent}, establishing it as the leading approach for short-wavelength, externally seeded FELs.

Operating short-wavelength seeded FELs at megahertz-class repetition rates is essential for photon-hungry experiments involving dilute samples and weak scattering signals \cite{macklin1996imaging,instruments3030047,Pedersoli_2013,Capotondi:ig5025,helfenstein2017coherent,bencivenga2015four,mochi2019lensless,kudilatt2020quantum}. The recent advent of superconducting linear accelerators (linacs) has enabled the delivery of high-brightness electron beams at megahertz frequencies, successfully supporting the operation of megahertz-class SASE-FELs \cite{ackermann2007operation,decking2020mhz}. However, achieving full longitudinal coherence at these repetition rates remains a formidable challenge for external seeding. Conventional seeding architectures require ultraviolet seed lasers with hundreds of megawatts of peak power to generate an energy modulation significantly larger than the intrinsic slice energy spread of the electron beam \cite{hemsing2014beam,feng2022coherent}. Given the average power limitations of existing laser systems, this stringent requirement restricts seed laser repetition rates to the kilohertz regime. Furthermore, while the standard EEHG scheme offers exceptional frequency up-conversion efficiency, its reliance on two independent seed laser pulses imposes stringent requirements on transverse spatial overlap and longitudinal timing synchronization. Such complexity can introduce additional timing instabilities during practical operation. Consequently, overcoming both the stringent peak-power requirements and the operational fragility of dual-seed configurations to achieve MHz-class, fully coherent X-ray FELs has remained a long-standing challenge \cite{schaper2021flexible,ackermann2020novel}.

Recently, the direct-amplification enabled harmonic generation (DEHG) technique \cite{wang2022high} was experimentally demonstrated to address this peak-power limitation \cite{qiFirst2025}. In its initial single-stage HGHG configuration, a weak ultraviolet seed with 0.75 µJ of pulse energy and megawatt-level peak power was amplified directly within a long modulator to drive saturated emission at the 7th harmonic. Crucially, this prior work confirmed that the amplified seed pulse preserves the full spatiotemporal coherence of the original seed, theoretically allowing it to be reused as the second seed in the EEHG configuration.

Here, we report the experimental extension of the DEHG concept into the beam-echo regime to achieve fully coherent, shorter-wavelength FELs. Originating from a single, low-power laser, the initial weak seed is amplified through its interaction with the electron beam in a primary long modulator, simultaneously imparting a sinusoidal-like energy modulation to the electrons. The amplified seed is then optically delayed and reused to modulate the electron beam a second time in a subsequent short modulator. Coupled with two dispersive sections, this process establishes a robust beam-echo effect, unlocking ultrahigh-harmonic generation without the need for a secondary, independent laser system.

Using this approach, we successfully demonstrated fully coherent FEL radiation amplified to saturation at the 16th harmonic of the seed laser, requiring an initial ultraviolet seed energy of only 0.4 µJ (2 MW peak power). This minimal energy requirement implies that a megahertz-repetition-rate EEHG-FEL could be driven by a standard commercial ultraviolet laser supplying just ~1 W of average power. Furthermore, coherent high-harmonic lasing was also realized up to the 20th and 30th harmonics, validating the inherent scalability of this technique to even shorter wavelengths. By dramatically reducing the seed power requirements and simplifying the operational architecture to a single seed, these experimental results establish a clear and practical pathway toward high-repetition-rate, fully coherent EUV and X-ray light sources.

\section*{Results}\label{sec2}

\begin{figure}[ht]
\centering
\includegraphics[width=1\textwidth]{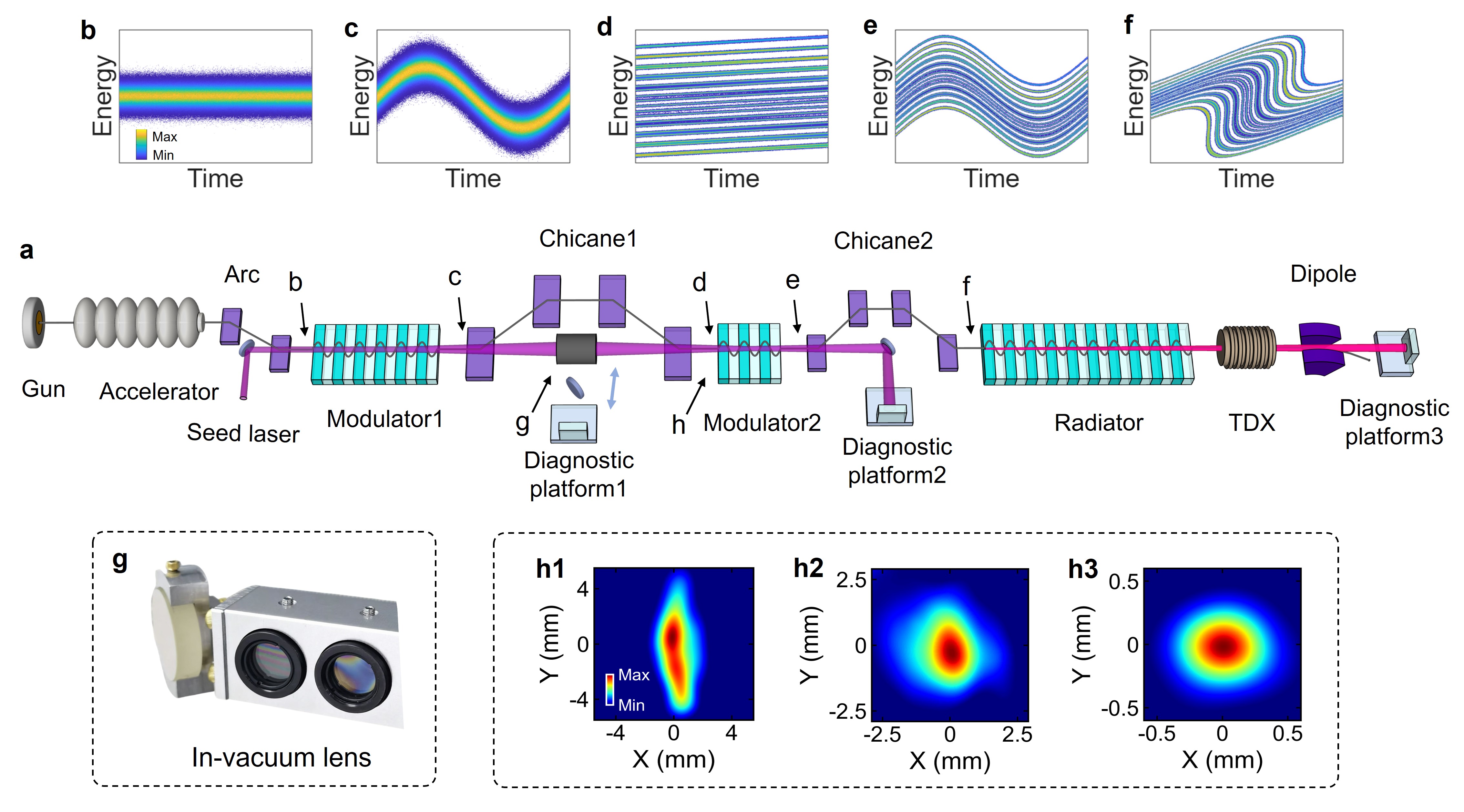}
\caption{\textbf{Experimental layout and beam dynamics of the DEHG scheme. a}, Schematic of the beamline at the SXFEL facility, illustrating the driver linac (photocathode gun and main accelerating structures), the ultraviolet seed laser system, the optical assembly for temporal delay and focusing, and the undulator configuration (comprising two modulators, two dispersive chicanes, and a downstream radiator). 
\textbf{b--f}, Simulated longitudinal phase-space evolution of the electron beam at critical stages: before the first modulator (\textbf{b}), after the first modulator (\textbf{c}), after the first chicane (\textbf{d}), after the second modulator (\textbf{e}), and after the second chicane (\textbf{f}). 
\textbf{g}, In-vacuum laser focusing and temporal delay optics assembly. This setup comprises two switchable lens barrels housing calcium fluoride (CaF\textsubscript{2}) lenses of different thicknesses. Both lenses maintain a 3.4 m focal length while introducing different temporal delays of 8.86 ps and 7.92 ps, respectively.
\textbf{h}, Measured transverse profiles of the seed at the entrance of the second modulator: the initial weak seed without amplification and focusing (\textbf{h1}), the amplified seed without focusing (\textbf{h2}), and the amplified seed after transverse focusing (\textbf{h3}).} \label{fig:setup}
\end{figure}
The experiment was performed at the Shanghai Soft X-ray Free-Electron Laser (SXFEL) facility \cite{liu2021sxfel,feng2022coherent}. The schematic layout of the experiment and the physical mechanism of the DEHG scheme are illustrated in Fig.~\ref{fig:setup}. The SXFEL linac delivered a high-brightness electron beam with a nominal energy of 890 MeV, a normalized projected emittance of 1.5 mm$\cdot$mrad, a bunch charge of 500 pC, a flat-topped longitudinal distribution ($\sim$1 ps duration), and a slice energy spread of $\sim$85 keV \cite{fengMeasurement2011}.


As shown in Fig.~\ref{fig:setup}, an initial weak ultraviolet seed laser (265.9 nm, $\sim$200 fs duration) with a pulse energy of only 0.4 µJ ($\sim$2 MW peak power) is injected into Modulator1. Through interaction with the electron beam in this 8-meter-long undulator, the weak seed is directly amplified to 1.6 µJ (reaching $\sim$8 MW peak power). Simultaneously, the electron beam acquires a sinusoidal-like energy modulation (Fig.~\ref{fig:setup}c), characterized by a dimensionless amplitude $A$ defined as the ratio of the induced energy modulation depth to the intrinsic slice energy spread of the electron beam. Crucially, despite this relatively modest seed laser power, the extended interaction length and the continuous, highly efficient coupling between the radiation and the electron beam during the FEL amplification process induce an $A$ exceeding 4. Achieving an equivalent modulation depth in a conventional short modulator (e.g., 1.28 m) would necessitate a seed laser with a peak power of approximately 70 MW (see Methods and Supplementary Fig.~\ref{fig:A_Z}). This highlights the effectiveness of the long-modulator in significantly reducing the peak-power requirements of the external seed laser. Furthermore, this amplified seed is uniquely characterized by a remarkable refinement of its transverse profile (Fig.~\ref{fig:setup}h1, h2). This "mode-cleaning" effect, driven by FEL gain-guiding dynamics within the undulator \cite{huang2007Review}, actively suppresses diffraction-induced divergence and purifies non-ideal initial transverse modes of the seed. The amplified radiation preserves a fully coherent ultraviolet bandwidth of $\sim$11.5 meV with high spectral stability (see Supplementary Fig.~\ref{fig:DE_spec}), serving as an ideal high-quality seed for the subsequent modulation stage.

Following Modulator1, the electron beam traverses a large dispersive section (Chicane1), which folds the energy modulation into fine-structured energy bands (Fig.~\ref{fig:setup}d) and introduces a temporal delay of several picoseconds. To reuse the amplified seed in Modulator2, it must be optically delayed and focused to maintain spatiotemporal overlap with the electron beam. We designed switchable, in-vacuum calcium fluoride (CaF$_2$) optics (Fig.~\ref{fig:setup}g) located in the middle of Chicane1. By exploiting the refractive index of the CaF$_2$ medium to reduce the group velocity of the laser pulse, this system is engineered to simultaneously achieve a transverse focal length of 3.4 m and introduce a precise group delay to compensate for the electron beam's transit through the chicane. To ensure broad operational flexibility, the optical platform features a dual-assembly design (housing both lenses and flat window plates of varying thicknesses), enabling step-tunable control over the refractive path length (see Methods). Notably, because Chicane1 intrinsically provides a continuous temporal-tuning range, precise spatiotemporal synchronization for the present experiment was elegantly achieved using only a single lens configuration (providing an 8.86 ps delay). After passing through this optic, the amplified seed is cleanly focused to a spot size of $\sim$610 µm (FWHM) before Modulator2 (Fig.~\ref{fig:setup}h3), establishing ideal conditions for the second electron-laser interaction. Upon entering Modulator2, the electron beam undergoes a second energy modulation (Fig.~\ref{fig:setup}e). A subsequent dispersion section (Chicane2) converts this complex energy phase space into a high-contrast density modulation (Fig.~\ref{fig:setup}f), successfully establishing a robust beam-echo effect capable of driving ultrahigh-harmonic generation in the downstream radiator. 

\begin{figure}[ht]
\centering
\includegraphics[width=1\textwidth]{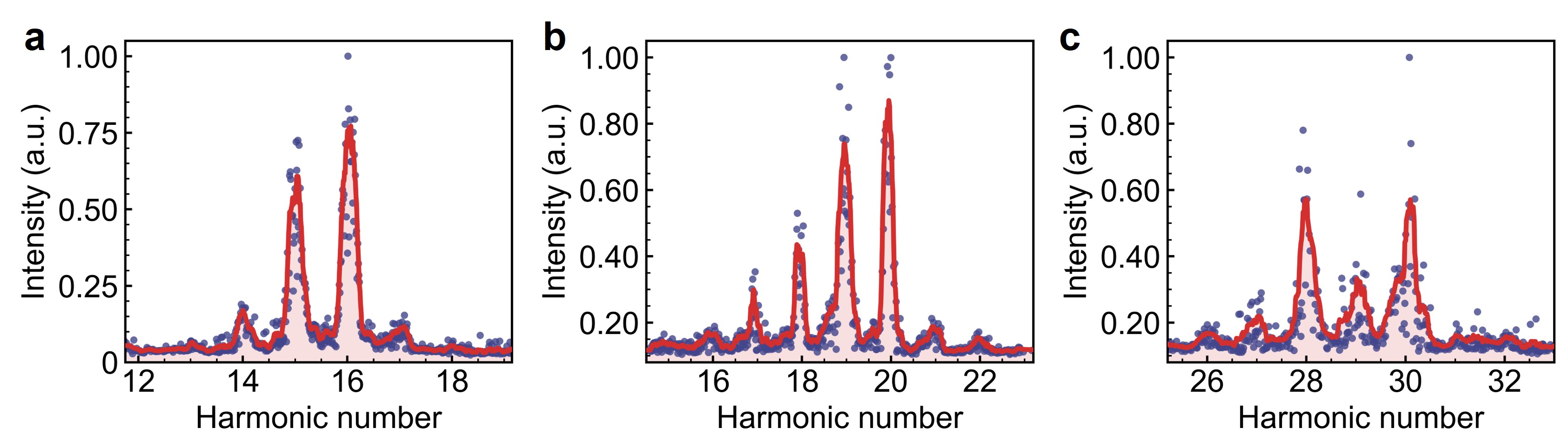}
\caption{
\textbf{Measured harmonic bunching distributions.}
\textbf{a}--\textbf{c}, Coherent radiation intensity profiles optimized for the 16th (\textbf{a}), 20th (\textbf{b}), and 30th (\textbf{c}) harmonics. Shifting the microbunching peak to these specific harmonic orders was primarily achieved by tuning the dispersion strength of Chicane2. The discrete intensity data points (blue markers) map the harmonic bunching distribution of the electron beam; these were acquired by systematically scanning the gap of the the first radiator segment to tune its resonant wavelength. The solid red curves represent the smoothed envelopes of the measured bunching distributions.} \label{fig:Harmonics}
\end{figure}

The efficiency of the beam-echo effect is dictated by the energy modulation amplitudes ($A_1$ and $A_2$) induced in Modulators 1 and 2, respectively. Using a coherent harmonic generation (CHG) based diagnostic method \cite{PhysRevAccelBeams.22.050703,PhysRevLett.115.114801}, we systematically optimized the beamline parameters. By measuring radiation intensities downstream of the radiator, we deduced an optimized $A_1 \approx 4$ and $A_2 \approx 1.6$ (detailed in Methods).
With the modulation amplitudes and chicane dispersion strengths ($R_{56}^1$ and $R_{56}^2$) properly tuned and synchronized, the DEHG configuration yielded exceptionally high-harmonic bunching distributions. To experimentally map these distributions, we systematically scanned the magnetic gap of the first radiator segment (while keeping the gaps of all downstream segments fully open) to sample the coherent radiation intensity across different resonant wavelengths. As illustrated in Fig.~\ref{fig:Harmonics}a, this measurement reveals a classic echo-enabled bunching signature strongly peaked at the 16th harmonic of the seed laser. We further demonstrated the inherent tunability of this scheme by optimizing the dispersion parameters to selectively shift the microbunching peak, driving strong coherent emission at the 20th and 30th harmonics (Fig.~\ref{fig:Harmonics}b, c) with an exceptionally high signal-to-noise ratio. The highest harmonic observed in this experiment (30th, 8.86 nm) was primarily limited by the minimal resonant undulator parameter ($K$) achievable at short wavelengths, rather than a fundamental limitation of the DEHG mechanism itself \cite{rebernik2019coherent}.

\begin{figure}[ht]
\centering
\includegraphics[width=1\textwidth]{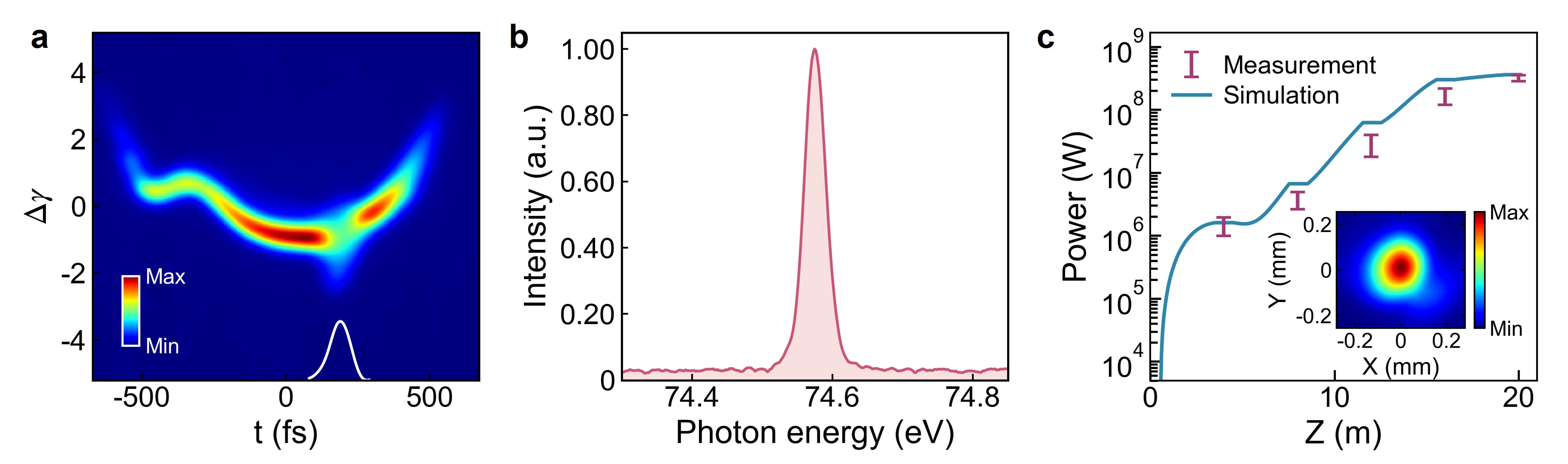}
\caption{
\textbf{Coherent, saturation lasing at the 16th harmonic.}
\textbf{a}, Measured longitudinal phase-space distribution of the electron beam post-lasing (bunch head to the left). The overlaid white profile represents the extracted energy loss, indicating an ultrashort FEL pulse duration of 82 fs (FWHM).
\textbf{b}, Single-shot spectral profile of the 16th harmonic ($\lambda \approx 16.62$~nm), exhibiting high spectral purity with a narrow relative bandwidth of $0.048\%$.
\textbf{c}, Measured FEL gain curve along the radiator, yielding a saturated output pulse energy of 28~µJ. The insets show the corresponding transverse intensity profiles, confirming the preservation of a near-ideal fundamental Gaussian mode throughout the amplification process.
}
\label{fig:16th gain}
\end{figure}
To comprehensively validate the performance of the DEHG FEL in the high-harmonic regime, the entire radiator section was configured to amplify the 16th harmonic (16.62 nm), taking advantage of the maximized undulator parameter $K$. As mapped by an X-band transverse deflecting cavity (Fig.~\ref{fig:16th gain}a), the seeded FEL lasing cleanly dominates the longitudinal phase space of the electron bunch, yielding an ultrashort pulse duration of $\sim$82 fs \cite{zeng2022online}. High-resolution single-shot spectra \cite{yang2024development} confirm fully coherent emission (Fig.~\ref{fig:16th gain}b), centered at 74.57 eV with a narrow 36 meV bandwidth that approaches 1.6 times the Fourier-transform limit. The measured pulse energy evolution along the radiator (Fig.~\ref{fig:16th gain}c) exhibits excellent agreement with three-dimensional Genesis 1.3 simulations \cite{reiche1999genesis}. This gain curve reveals a robust amplification process with a short gain length of $\sim$1.3 m, driving the FEL to saturation. The maximum output pulse energy reached 28 µJ, corresponding to a peak power of $\sim$350 MW with a 9.8\% rms energy stability, while the transverse intensity profile (Fig.~\ref{fig:16th gain}c, inset) maintained a fundamental Gaussian distribution with a spot size of $\sim$150 µm (FWHM).

\begin{figure}[ht]
\centering
\includegraphics[width=1\textwidth]{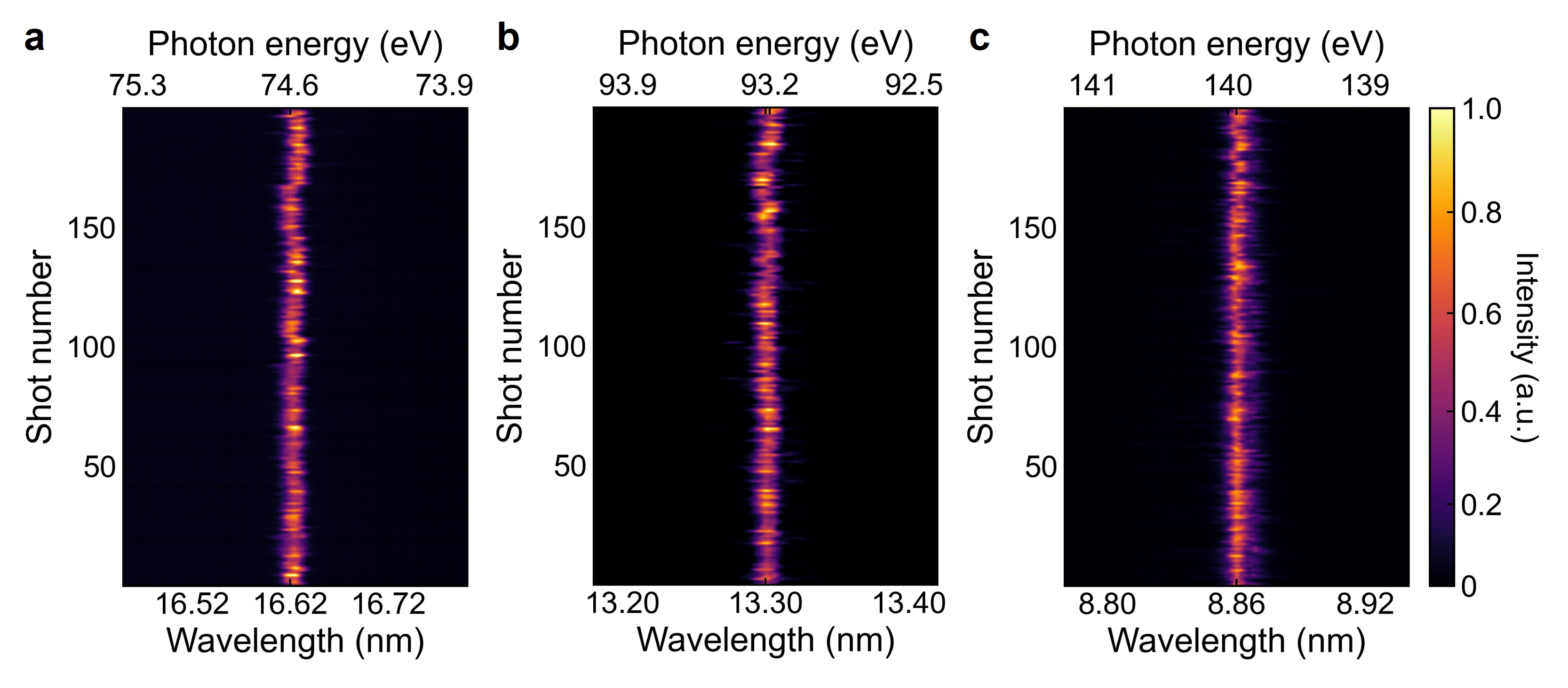}
\caption{
\textbf{Single-shot spectral stability at high harmonics.} 
\textbf{a}--\textbf{c}, Overlaid spectra from 200 consecutive single-shot measurements for the 16th (\textbf{a}, 16.62~nm), 20th (\textbf{b}, 13.30~nm), and 30th (\textbf{c}, 8.86~nm) harmonics. The corresponding relative r.m.s. central wavelength stabilities are $2 \times 10^{-4}$, $1.4 \times 10^{-4}$, and $1.6 \times 10^{-4}$, respectively.
}
\label{fig:spec}
\end{figure}

Building upon the saturated 16th harmonic, we extended the DEHG FEL operation to the 20th and 30th harmonics. This was accomplished by fine-tuning the dispersion strength of Chicane2 ($R_{56}^2$) and the radiator's resonant wavelength, alongside optimizing the seed amplification in Modulator1 to slightly increase the modulation amplitudes ($A_1$ and $A_2$) for enhanced high-harmonic bunching. Consequently, the emission yielded pulse energies of approximately 5.5 µJ and 0.1 µJ at radiator $K$ parameters of 1.84 and 1.25, respectively. While the 20th harmonic exhibited strong exponential growth, it did not reach full saturation within the limited interaction length of the five-segment radiator. For the 30th harmonic, the radiation power was primarily constrained by the small radiator $K$ value required for such short wavelengths, which inherently weakens the FEL coupling. Despite these power constraints, the emission intensity remained sufficient to provide a high signal-to-noise ratio for downstream diagnostics. Using the X-ray spectrometer on Diagnostic Platform3 (see Methods), we recorded 200 consecutive single-shot spectra (Fig.~\ref{fig:spec}) to evaluate the output stability. The system achieved exceptional relative central wavelength stabilities of $2 \times 10^{-4}$, $1.4 \times 10^{-4}$, and $1.6 \times 10^{-4}$ for the 16th, 20th, and 30th harmonics, respectively. These measurements clearly demonstrate the remarkable operational robustness of this single-seed setup, as the radiation preserved highly consistent and coherent profiles across all targeted harmonics.

\section*{Discussion}\label{sec12}
In summary, we have experimentally demonstrated a fully coherent, short-wavelength FEL driven by a single sub-microjoule seed. Eliminating the hundred-megawatt peak-power requirements of traditional external seeding and the complexities of dual-laser synchronization, the DEHG method provides a highly stable and simple approach for driving high-repetition-rate EEHG-FELs. Crucially, this in-situ amplification process serves two functions: the continuous radiation-beam coupling induces a deep energy modulation, while FEL gain-guiding dynamics naturally purify the seed laser through a "mode-cleaning" effect. With an initial ultraviolet seed of only 0.4 µJ, we achieved saturated FEL lasing at the 16th harmonic and robust coherent emission down to the 30th harmonic (8.86 nm). By bringing the energy demand down to this sub-microjoule level, this approach makes it feasible to drive megahertz-rate, short-wavelength FELs using standard commercial ultraviolet lasers operating at average powers of just $\sim$1 W.

Looking forward, the performance of this single-seed scheme can be further optimized. Theoretically, the initial seed power could be reduced even further by employing a longer modulator (Modulator1), provided the peak power of the seed remains sufficient to dominate the intrinsic shot noise of the electron beam. Additionally, employing a longer Modulator2 would provide a broader tuning range for $A_2$, offering greater flexibility to precisely optimize the bunching efficiency for specific target harmonics. On the optical engineering front, the setup adopted in this proof-of-principle experiment relies on a relatively thin in-vacuum lens optimized specifically for the 16th harmonic. The most straightforward way to reach higher harmonics is simply inserting a thicker lens to introduce a larger optical delay. This allows the system to operate with a stronger dispersion in the first chicane (characterized by $B_1$ or $R_{56}^1$), pushing the FEL emission to much shorter wavelengths while keeping the modulation amplitudes ($A_1$ and $A_2$) constant. Taking this a step further, implementing a variable delay system would completely streamline the synchronization process, maximizing the  flexibility of this method to cover a much broader spectral range.

Beyond these system-level enhancements, extending fully coherent FEL operation toward the deep X-ray regime remains limited by the initial seed laser wavelength. However, recent advances in high-order harmonic generation (HHG) in noble gases, which now deliver MW-level peak power pulses in the EUV range \cite{klas2021,shirozhan2024,saule2019}, offer a natural solution\cite{lambert2008injection,dunning2011design,giannessi2006free,ackermann2013generation}. Because the DEHG mechanism lowers the energy threshold for the seed and corrects spatial flaws through the "mode-cleaning" effect, these weak HHG pulses become viable drivers. Combining the short wavelengths of HHG with the high amplification of the DEHG could push coherent emission into the water window and tender X-ray regimes. Ultimately, this capability would enable time-resolved studies of quantum materials and minimally invasive imaging of biological systems \cite{liAttosecondpump2024,kordel2020,jordan2020,reinhard2026}.

\section*{Methods}
\subsection*{Machine setup}
The experimental layout at the SXFEL facility is illustrated in Fig.~\ref{fig:setup}. An 890 MeV electron beam is delivered from the linear accelerator (linac) into the seeding undulator (SUD) line via the arc switchyard. A single weak ultraviolet seed laser ($\sim$265.9 nm, generated via the third harmonic of a Ti:sapphire laser) is injected into the beamline just downstream of the arc. The primary modulator (Modulator1) consists of two 4-meter-long undulator segments with a period length of 68 mm, tuned to an undulator parameter $K \approx 6.74$. Chicane1 is a strong dispersive section with a continuously tunable momentum compaction factor ($R_{56}^1$) ranging from 0 to 10 mm. During the experiment, it was typically set to several millimeters to fold the energy modulation induced in Modulator1 into fine-structured energy bands.

An in-vacuum calcium fluoride (CaF$2$) lens assembly focuses and delays the amplified seed pulse. This allows the same seed to be reused to modulate the electron beam a second time in Modulator2, a 1.28-meter-long undulator section with an 80 mm period and a $K$ value of 6.19. Chicane2, featuring a tunable $R_{56}^2$ from 0 to 2 mm, provides the small dispersion necessary to convert this second energy modulation into a density modulation. This process drives strong ultrahigh-harmonic bunching within the electron beam, seeding the coherent short-wavelength FEL emission in the subsequent radiator section. The radiator itself comprises five 3-meter-long undulator segments with a 30 mm period length.

Following the radiator, the spent electron beam passes through an X-band transverse deflecting cavity (TDX) before a bending magnet (dipole) directs it to the beam dump. This configuration maps the longitudinal phase space of the electron beam onto an interceptive profile monitor. Simultaneously, the emitted FEL pulses propagate down the photon beamline for diagnostics and downstream experiments.

The beamline incorporates three primary diagnostic platforms. Diagnostic Platform1 (Supplementary Fig.~\ref{fig:Diagnostics}) is primarily dedicated to characterizing the seed laser both before and after its initial amplification. It is equipped with a fast photodiode (PD) for temporal synchronization with the electron beam, a pyroelectric sensor for pulse energy measurements, a CCD-coupled fluorescent screen for transverse profile imaging, and an ultraviolet spectrometer for spectral analysis (see Supplementary Fig.~\ref{fig:DE_spec} for the measured spectra). Diagnostic Platform2 mirrors this configuration; installed immediately after the second modulation stage, it serves to monitor the status of the reused seed laser in real-time. Finally, Diagnostic Platform3 is located at the downstream photon beamline. It characterizes the final FEL spot size, pulse energy, and spectrum using a profile monitor, a photodiode energy sensor, and a high-resolution X-ray spectrometer \cite{yang2024development}, respectively.

\subsection*{In-vacuum lens}
The in-vacuum lens system (Fig. \ref{fig:setup}g and Supplementary Fig. \ref{fig:cavity}) was custom-designed to simultaneously focus the diverging amplified seed laser and temporally synchronize it with the electron beam.

For spatial focusing, the lens is positioned to image the laser waist onto the center of Modulator2. Based on standard Gaussian beam propagation within the SXFEL beamline layout, we required a focal length of $f = 3.4$ m. The lens assembly is located 5.2 m downstream from the exit of Modulator1 and 6.3 m upstream from the center of Modulator2. This geometry focuses the transverse spot to a full width at half maximum (FWHM) of $<500$ µm. This spot size intentionally overfills the electron beam (transverse size $\sim$70 µm) while maintaining a sufficient optical energy density for the second-stage modulation. To account for manufacturing tolerances, all lenses were characterized offline, and only those matching the exact 3.4 m focal length were installed. Precise transverse spatial overlap between the laser and the electron beam is achieved using a motorized two-dimensional translation mechanism integrated into the assembly (Supplementary Fig. \ref{fig:cavity}).

Beyond spatial focusing, the assembly provides a step-tunable group delay for temporal synchronization. This delay arises from the refractive index of the calcium fluoride ($\text{CaF}_2$) medium, which reduces the group velocity of the laser pulse relative to vacuum, compensating for the electron beam's longer transit path through Chicane1. To allow adjustable control over this optical delay, the vacuum chamber houses a dual-module mechanism. The first module contains switchable lenses of varying center thicknesses (all maintaining $f = 3.4$ m), while a second independent module holds flat $\text{CaF}_2$ window plates of different thicknesses to provide up to 1 ps of supplementary delay. Motorized actuators enable switching between these optical elements.

The optical delay of each element was calibrated offline using a spectral interference technique, where exact delay times were extracted directly from the interference fringes (Supplementary Fig. \ref{fig:calibration}). These measurements confirmed discrete temporal delays of 8.86 ps and 7.92 ps for two distinct lens configurations. In practice, while the dual-assembly setup provides coarse step-tuning for the delay, the continuous dispersion adjustment of Chicane1 ($R_{56}^1$) handles the fine synchronization. For the experiments reported here, synchronization was achieved using a single lens configuration (8.86 ps delay) without supplementary flat window plates. This combination of coarse optical delay and fine magnetic tuning minimizes operational complexity while maintaining the target EEHG beam parameters.

\subsection*{Experimental tuning and optimization}
EEHG is well known for its exceptional frequency-up-conversion efficiency using only modest energy modulation strengths $A_1$ and $A_2$~\cite{stupakov2009,xiang2009echo,zhao2012first,rebernik2019coherent}. Consequently, the tuning and optimization of these two parameters were of essential importance to this experiment. In seeded FELs, the standard procedure for determining the energy modulation strength relies on the coherent harmonic generation (CHG)-based diagnostic method. The intensity of the CHG radiation is proportional to the square of the high harmonic bunching factor $b_n$, which is expressed as 
\begin{equation*}
    b_n = e^{-\frac{1}{2} n^2 B^2} \left| J_n(-nAB) \right| 
\end{equation*}
where $A$ is the dimensionless energy modulation strength and $B$ is the dimensionless dispersive strength in relation to the Chicane $R_{56}$.
The energy modulation amplitude $A$ can be deduced by monitoring the CHG radiation intensity while scanning $R_{56}$ around its optimal value.

In our experiment, the CHG radiation was produced by the first radiator segment, which is a dedicated diagnostic undulator employing dual-period magnetic arrays for different CHG harmonic numbers. 
The radiation intensities were detected by a photodiode energy sensor located at Diagnostic Platform3. 
To characterize the energy modulation ($A_1$) introduced in Modulator1, the dispersion $R_{56}$ was optimized to 130~µm, yielding an inferred modulation amplitude of $A_1 \approx 4$. We also measured the CHG radiation intensities at various harmonic numbers (Supplementary Fig.~\ref{fig:A1A2}a). Coherent radiation signals were observed up to the 7th harmonic, exhibiting excellent agreement with theoretical predictions. Numerical simulations indicate that achieving this equivalent modulation amplitude within a conventional short undulator (e.g., Modulator2) would necessitate a peak seed-laser power of approximately 70~MW (Supplementary Fig.~\ref{fig:A_Z}), By contrast, the DEHG mechanism can achieve this target using only a 2~MW seed over a 6-m interaction length, with the full 8-m length of Modulator1 utilized experimentally to ensure a robust operational margin.

After $A_1$ was optimized and determined, Chicane1 was set to be a large $R_{56}^1$ to fold the phase space and form the fine energy bands. The amplified seed laser was focused and delayed by the in-vacuum lens with a focal length of 3.4 m and a precise time delay of 8.86 ps, and then transported to Modulator2 to introduce a second energy modulation $A_2$. To tune and optimize $A_2$ accurately, we employed a fresh-bunch technique in the experiment. $R_{56}^1$ of Chicane1 is chosen to be 5.57 mm, corresponding to a 9.28 ps time delay of the electron beam, which is approximately 420 fs more than the laser delay, so that the amplified seed laser can interact with a fresh part of the electron beam. By tuning the resonant interaction between the seed laser and the fresh electron bunch, $A_2$ was determined to be about 1.6 with the optimized $R_{56}^2$ being 340 µm. The CHG radiation intensities for different harmonic numbers are shown in Supplementary Fig.~\ref{fig:A1A2}b. After $A_2$ was optimized in the fresh part of the electron beam, $R_{56}^1$ was set back to 5.32 mm which corresponds to the exact same delay with the optical lens, so that the two-stage energy modulations were synchronized together, leading to the formation of the beam echo effect in the electron beam.                 

\subsection*{Simulation}
Numerical simulations were performed using the three-dimensional FEL code Genesis 1.3 \cite{reiche1999genesis} to benchmark the experimental results. Using the experimentally determined beam and laser parameters as initial inputs, we simulated the complete EEHG beamline, from the entrance of Modulator1 to the exit of the radiator. To accurately reproduce the measured FEL performance, the simulation tuning procedure directly mirrored the experiment. Specifically, we scanned the effective peak power of the reused seed laser in Modulator2, the dispersion strength of Chicane2 ($R_{56}^2$), and the radiator resonance to optimize phase-space folding and maximize high-harmonic bunching.

\section*{Data availability}

The data that support the findings of this study are presented in the article and the Supplementary Information. Other relevant data and findings of this study are available from the corresponding authors upon reasonable request.

\section*{Acknowledgements}

This work was supported by the SXFEL facility, the National Natural Science Foundation of China under grant numbers 12405363 (Z.Q.), 12435011 (C.F.) and 12475322 (T.L.), CAS Project for Young Scientists in Basic Research under grant number YSBR-115 (C.F.) and YSBR-091 (YX.F.), Shanghai Municipal Science and Technology Major Project (C.F.) and Innovation Program of Shanghai Advanced Research Institute, CAS, under grant number 2025CP003 (Z.Q.), the Youth Innovation Promotion Association CAS (T.L.) and the Oriental Talent Young Scientist Program QNZH2024036 (T.L.).

\section*{Author contributions}

C.F. and Z.Z conceived the original idea and proposed the experiment at SXFEL, Z.Q., L.N., Xingtao Wang, W.Y.and C.F. designed the experiments. The experiments were conducted by T. Liu, Z.Q., L.N. and C.F. with help from Z.W., K,Z, Z.G., H.Y., N.H., H.L., S.C., J.L., Y.X., L.T., Xiaofan Wang. D.X, B. Liu and with software and hardware support by C.Y., Y.W., F.G., Y.L., J.C., H.Z., X.L., L.F., Y.Z., J.X., L.S., W.Z., Y. W., T.Lan. X.Z., and B. Li. The simulations and data analysis were performed by L.N. and Z.Q. The paper was written by L.N., Z.Q. and C.F. with contributions from T. Liu. Management and oversight of the project was provided by C.F. and Z.Z.

\backmatter

\begin{appendices}

\section{Additional information}\label{secA1}

\begin{figure}[!hp]
\centering
\includegraphics[width=0.8\textwidth]{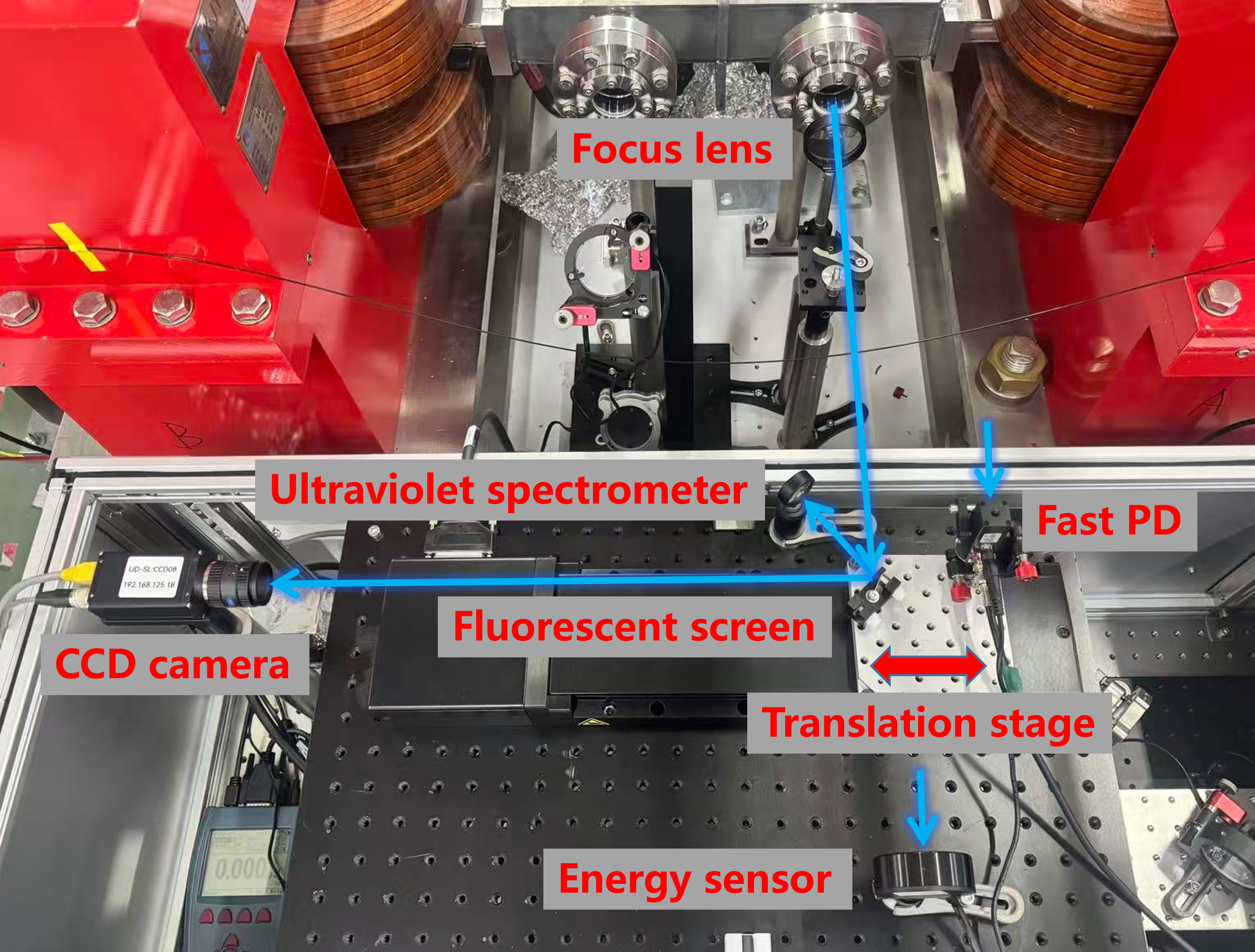}
\caption{
\textbf{Seed laser diagnostic platform.} 
Photograph of the online diagnostic setup (Diagnostic Platform1) used to characterize the seed laser. The platform integrates a fast photodiode, a pyroelectric energy sensor, a CCD-coupled fluorescent screen, and an ultraviolet spectrometer to monitor the optical pulses both before and after amplification.
}
\label{fig:Diagnostics}
\end{figure}

\begin{figure}[!hp]
\centering
\includegraphics[width=0.8\textwidth]{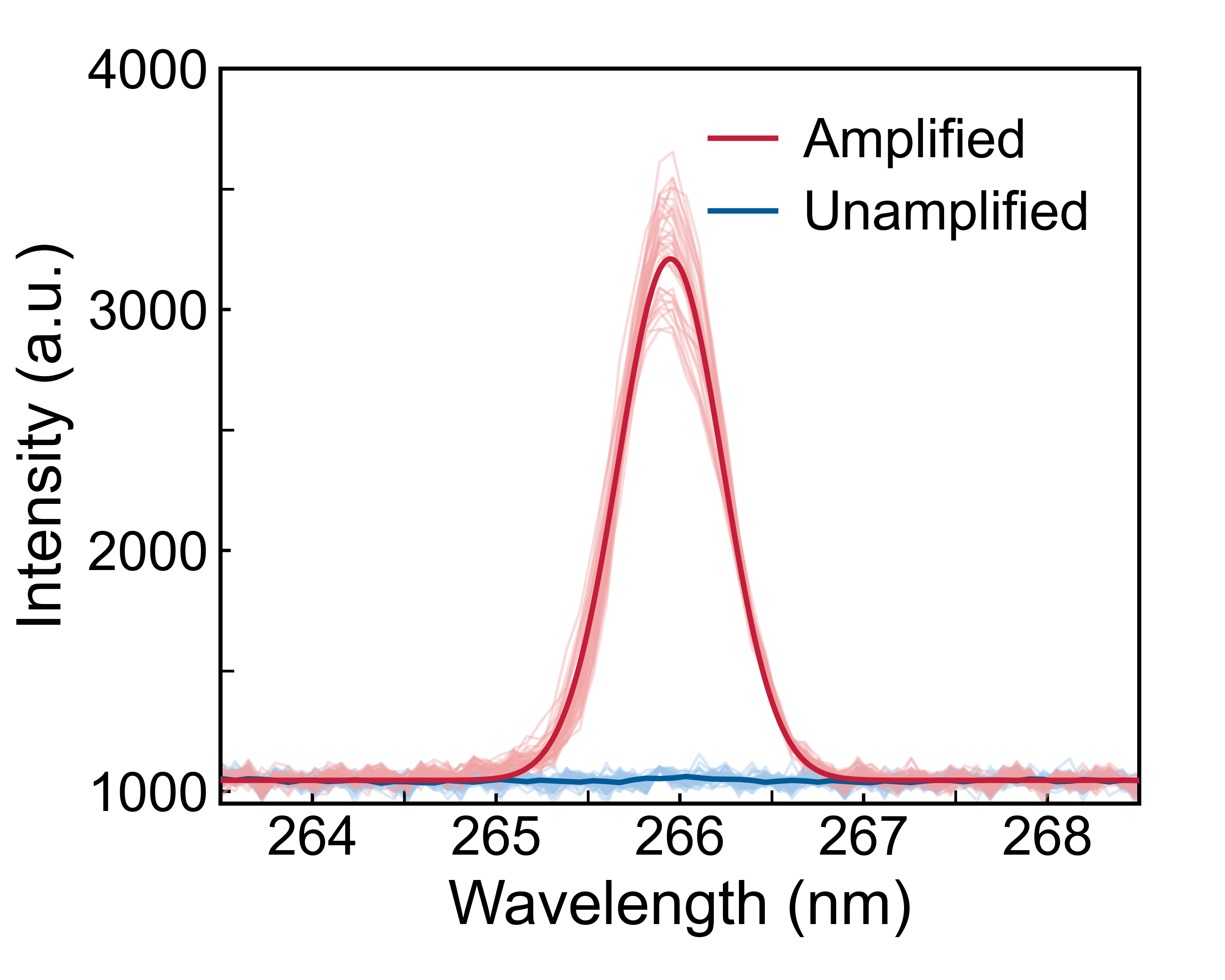}
\caption{
\textbf{Seed laser spectrum after initial amplification.} 
Measured spectrum of the ultraviolet seed laser after interacting with the electron beam in Modulator1. The amplified seed preserves a fully coherent bandwidth of $\sim$11.5~meV, confirming its suitability as a high-quality driver for the subsequent modulation stage.
}
\label{fig:DE_spec}
\end{figure}

\begin{figure}[!hp]
\centering
\includegraphics[width=0.8\textwidth]{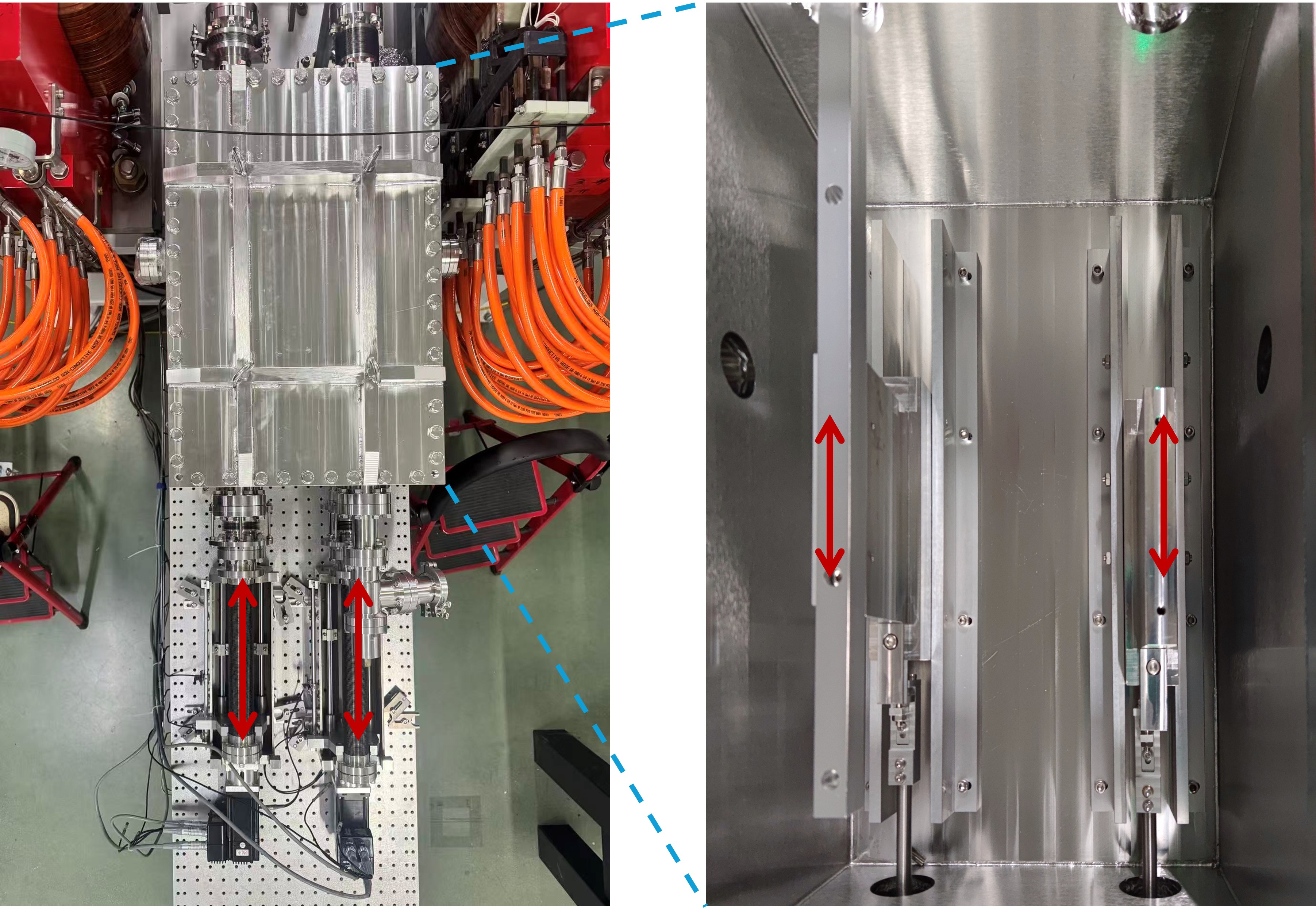}
\caption{
\textbf{Mechanical design of the in-vacuum lens assembly.} 
\textbf{Left}, external view of the vacuum chamber housing the optical elements. \textbf{Right}, internal layout detailing the motorized dual-module mechanism. This assembly enables both precise two-dimensional transverse alignment for spatial focusing and the switching of CaF$_2$ lenses and flat plates to provide step-tunable optical delays for temporal synchronization.
}
\label{fig:cavity}
\end{figure}

\begin{figure}[!hp]
\centering
\includegraphics[width=0.75\textwidth]{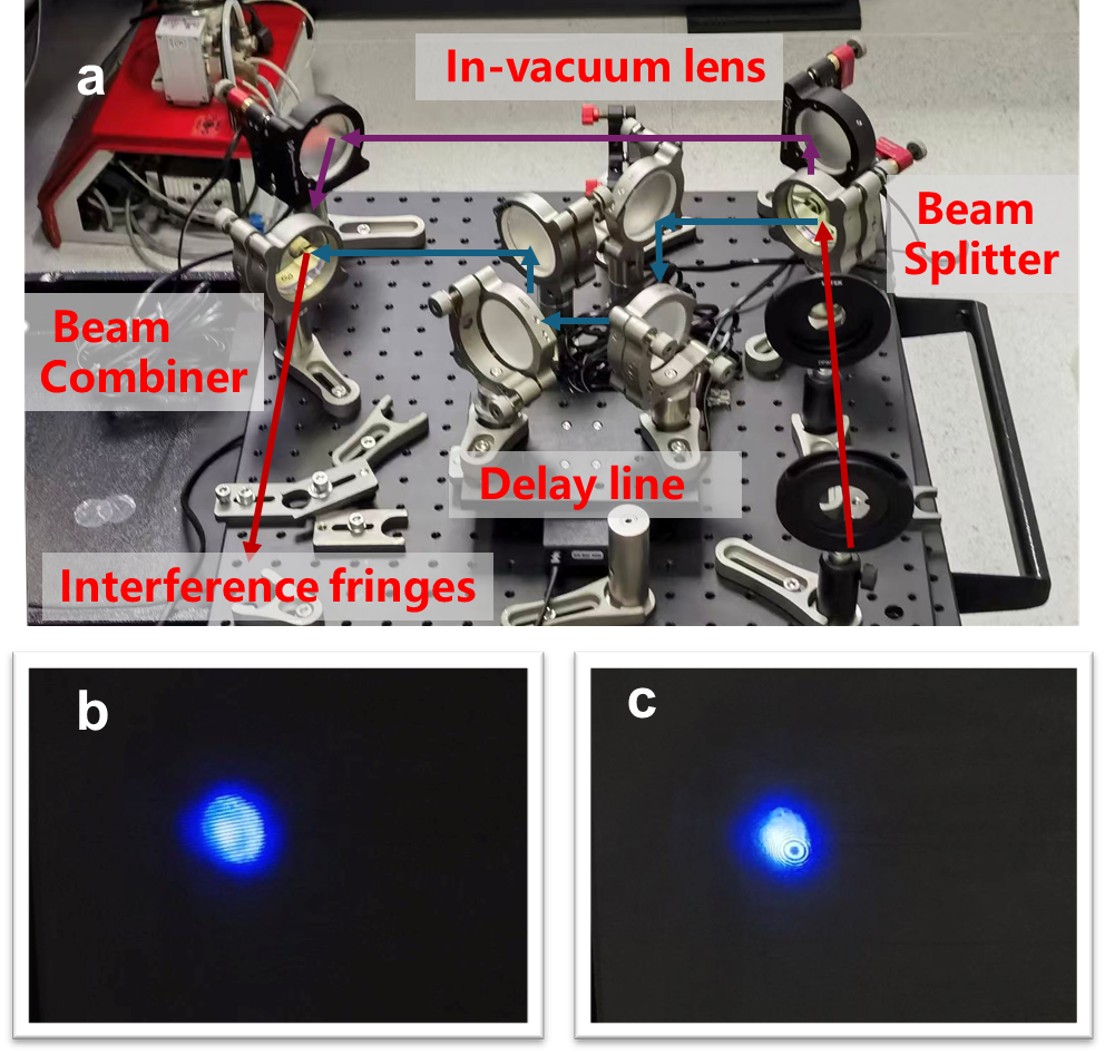}
\caption{
\textbf{Offline calibration of the optical group delay.} 
\textbf{a}, Schematic of the diagnostic optical path utilizing a spectral interference technique. \textbf{b}, \textbf{c}, Measured spectral interference fringes for the flat CaF$_2$ window (\textbf{b}) and the CaF$_2$ lens (\textbf{c}) housed within the lens assembly. These interference patterns are used to precisely extract the exact temporal delays (e.g., 8.86~ps for the lens) introduced by each optical element.
}
\label{fig:calibration}
\end{figure}

\begin{figure}[!hp]
\centering
\includegraphics[width=0.85\textwidth]{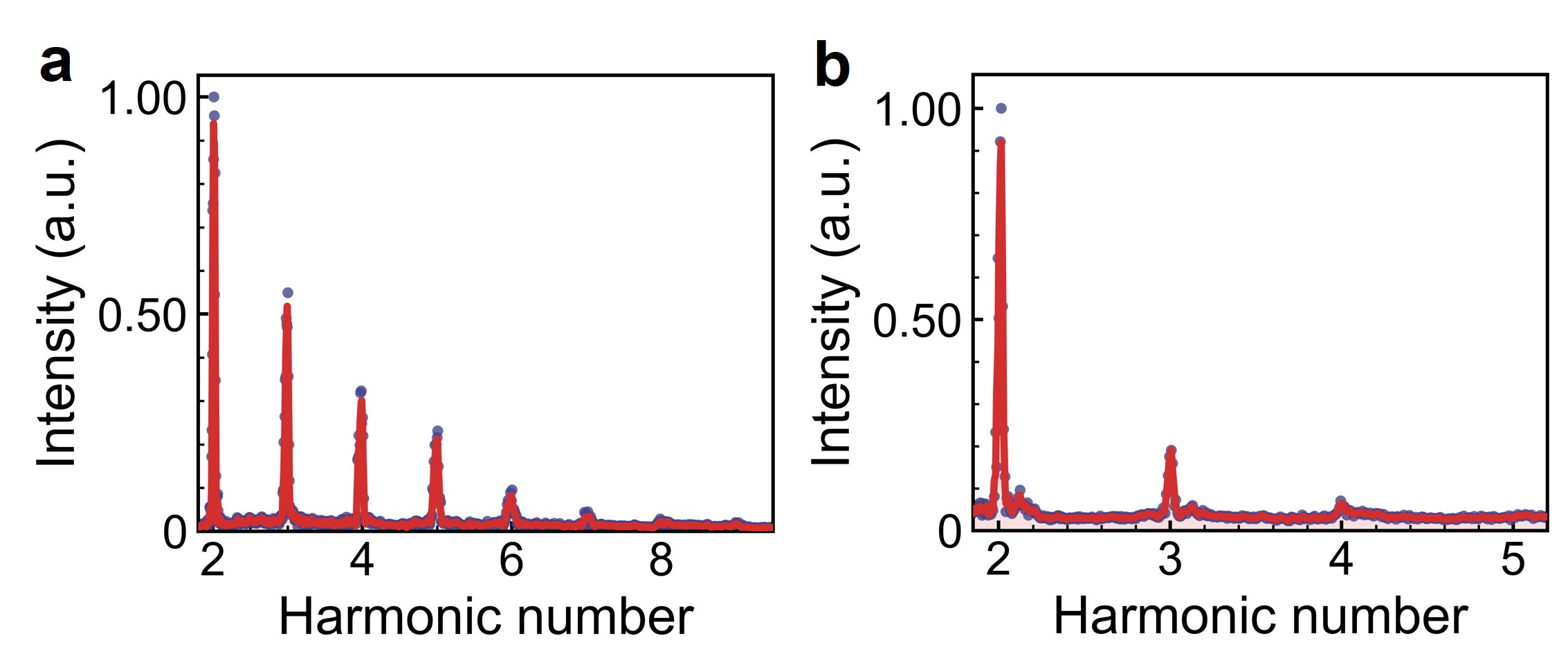}
\caption{
\textbf{Coherent harmonic radiation from individual modulation stages.} 
\textbf{a}, \textbf{b}, Measured radiation intensities at various harmonic orders when operating only the first (\textbf{a}) or second (\textbf{b}) modulator--chicane stage. Blue markers represent the experimental data, and red curves indicate the fitting results. These distributions are used to accurately deduce the induced energy modulation amplitudes, yielding $A_1 \approx 4$ (optimized at $R_{56}^{(1)} = 0.13~\mathrm{mm}$) and $A_2 \approx 1.6$ (optimized at $R_{56}^{(2)} = 0.34~\mathrm{mm}$). $A$ denotes the dimensionless energy modulation amplitude, and $R_{56}$ is the chicane dispersion strength.
}
\label{fig:A1A2}
\end{figure}

\begin{figure}[!hp]
\centering
\includegraphics[width=1\textwidth]{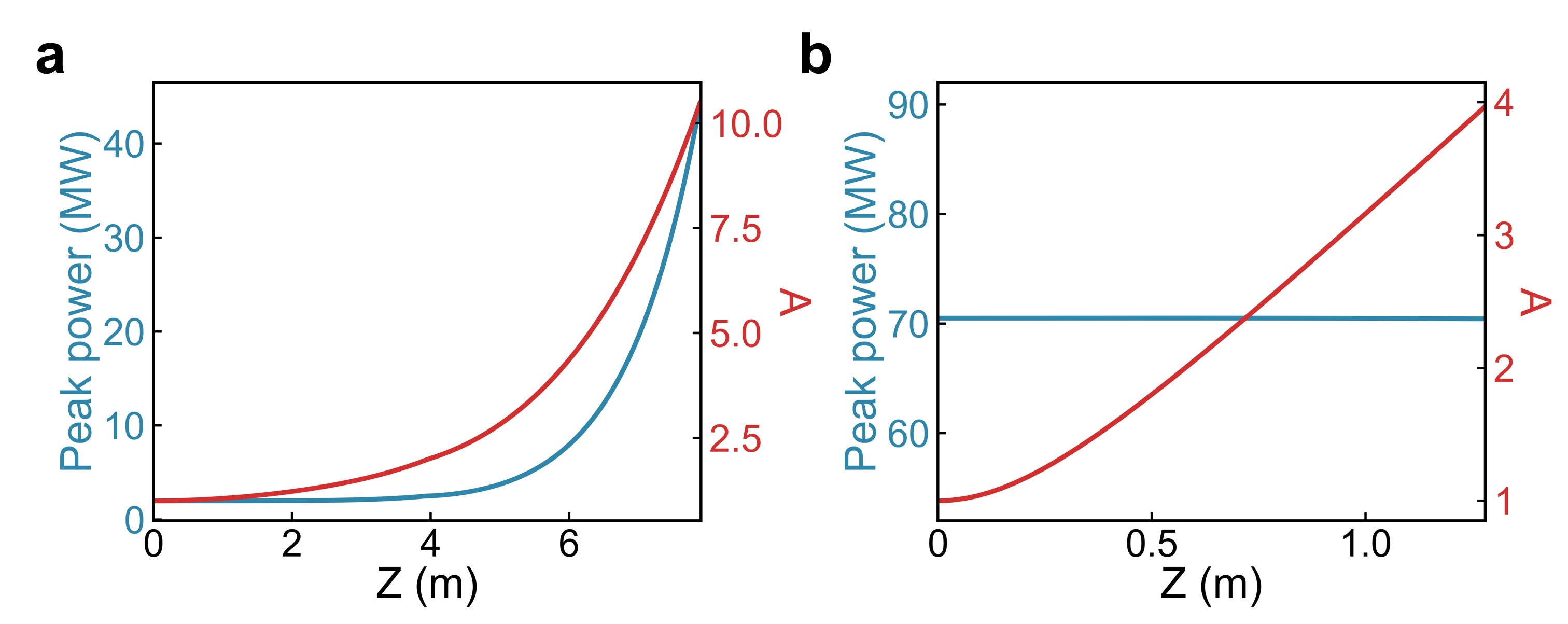}
\caption{
\textbf{Seed laser amplification and energy modulation dynamics.}
\textbf{a}, \textbf{b}, Simulated evolution of the laser peak power (blue curves, left axes) and the dimensionless energy modulation amplitude, $A$ (red curves, right axes). The DEHG process in the long Modulator1 driven by a 2~MW initial seed (\textbf{a}) is compared with a conventional short undulator driven by a 70~MW seed (\textbf{b}). This comparison demonstrates that reaching the target modulation depth of $A \approx 4$ in a standard short modulator necessitates a significantly higher initial peak power.
}
\label{fig:A_Z}
\end{figure}




\end{appendices}

\clearpage 

\bibliography{sn-bibliography}

\end{document}